\documentstyle[preprint,aps]{revtex}
\begin{document}
\baselineskip 0.7cm
\input{psfig.sty}
\title{\hfill hep-ph/9610440 (revised March 7 1997)\\ 
Breaking Flavour Symmetry Spontaneously.}
\author{Nils A. T\"ornqvist}
\address{Research Institute for High Energy Physics, SEFT
POB 9, FIN--00014, University of Helsinki, Finland}
\date{7 March  1997}                 
\maketitle
\begin{abstract}
A new mechanism for spontaneous breaking of flavour symmetry is demonstrated.
An exactly flavour symmetric model with degenerate bare nonets and 
with sufficiently strong tri-linear meson couplings is shown
 to lead to self-consistency equations which 
are  unstable. Instead there exists a stable solution,
 which break flavour symmetry  spontaneously in the mass spectrum.
 For a $C$-degenerate meson spectrum 
the stable mass spectrum  obeys the Okubo-Zweig-Iizuka 
(OZI) rule and the approximate  
equal spacing rule. 
\\
\noindent Pacs numbers:12.39.Ki, 11.30.Hv, 11.30.Qc, 12.15.Ff
\vskip 0.20cm 
\end{abstract}

{\it 1. Introduction and the NJL model.}                                    
Any solution of the hadron mass spectrum, consistent with QCD, must
satisfy the constraints of: ({\it i}) unitarity, ({\it ii}) analyticity, and
in the massless quark limit also
({\it iii}) chiral symmetry and ({\it iv})  flavour symmetry.

Conventionally one breaks ({\it iii-iv}) by adding effective
non-degenerate quark masses to the Lagrangian,  wherby
the pseudoscalars obtain (small)  masses
and the degeneracy of all flavour  multiplets is split. The gluon anomaly
 contributes to the flavourless sector in breaking 
the symmetry to approximate $SU3_f$, giving especially 
the $\eta$ mesons part of their masses. 
Most of the chiral quark masses are assumed to come from a
short distance regime, where  weak interactions are relevant.
Phenomenologically fermion masses can be generated by suitable {\it ad hoc}
couplings to a Higgs field. Here I shall discuss an alternative way of 
breaking the flavour symmetry. 

From deep inelastic scattering, including the spin problem of the
proton, we know that mesons and
baryons are not simple $q\bar q$ and $qqq$ quark model bound states, 
but have, in addition to the valence quarks, large 
components in the form of a  multiquark $q\bar q$ sea and gluons. 
Thus a constituent quark of the naive quark model must, in fact, be composed 
of many quarks, antiquarks and gluons. This requires a multichannel formalism,
which couples $q\bar q$ and multiquark states.  

Recently it was shown that one can obtain a good understanding of
the light scalar meson nonet~\cite{NAT} when 
one uses the constraints ({\it i-iv}) above provided
 one includes in a coupled channel framework all light two
pseudoscalar thresholds  
with flavour symmetry broken mainly by  the pseudoscalar masses.
Similar models for other multiplets also improve our understanding
of the hadron spectrum, even for heavy quarks. One good example is
 the anomalously large $\Upsilon (5)$-$\Upsilon (4)$ 
mass splitting~\cite{NATup}, 
where single channel potential models 
clearly fail, because of the opening of the $B\bar B$, $B\bar B^*$ etc.
thresholds.

In such models one needs a smaller
bare $m_s-m_d$ quark mass term than usual, and this becomes smaller 
the more hadronic thresholds are included. With sufficient 
number of higher thresholds (obeying the empirically approximate
 equal spacing rule) 
 the $SU3_f$ breaking usually attributed to $m_s-m_d$ 
could be  generated by hadronic loops. This 
opens up the possibility that the bare $m_s-m_d$ could be put equal zero,
and that flavour symmetry breaking could be generated spontaneously,
through the self-consistency equations for hadron propagators, which
include hadronic thresholds. 

 The well known Nambu--Jona-Lasinio (NJL) model~\cite{nambu} 
is a standard reference model for for chiral symmetry breaking and for
dynamical symmetry breaking in general. Although I shall not use this 
model here the instability mechanism in the original NJL model has similarities
with the mechanism I shall discuss, and therefore I recall here some well
known results from the NJL model. 
For  reviews see Ref.~\cite{NJL}. 
With a small chiral mass ($\mu$) the well known gap equation for the 
quark constituent mass ($M$) is
\begin{equation}
\frac M \Lambda =\frac \mu \Lambda +
\frac M \Lambda  \frac{\alpha_{NJL}\pi}{4}
{\rm F}(0,\frac {M^2} {\Lambda^2},\frac{M^2}{\Lambda^2},1) \ . \label{gap3}
\end{equation}
Here $\Lambda$ is the cutoff and $\alpha_{NJL}$ measures the strength of the
four fermion interaction. The function F takes into account the dependence
on the fermion mass in the tadpole-like loop. It will be defined below in a
dispersive framework.

Eq.~(\ref{gap3})  has for $\alpha_{NJL} >1,\ \mu =0$, 
apart from the trivial, unstable solution
$M=0$, a massive and stable solution. 
This is most easily seen by plotting the
left side of the equation, $M_{out}/\Lambda$,
against $M_{in}/\Lambda$ of the right side as in Fig.~1. For
$\alpha_{NJL} >1$ the slope of the curve  at the origin is $>1$. Then starting
at any value for $M_{in}/\Lambda$ and iterating the equation one ends up at
the massive solution as shown in Fig.~1.   On the other hand if
$\alpha_{NJL} <1$ one obtains the dashed curve and the trivial massless
solution  is the stable one.

At the same time as the fermions obtain mass, poles are generated in the
pseudoscalar and scalar $q\bar q$ channels. The pion is massless 
as required by chiral symmetry while the $\sigma$ has a mass of $2M$. 
In generalizations one expects a massless $0^{-+}$ multiplet and
massive but degenerate multiplets of other $J^{PC}$ when the $SU(N_f)$
remains unbroken.    In the following I shall  not rely further  on
 the NJL model. 

{\it 2. A flavour symmetric scalar model.} 
We now construct a meson model, 
 where  all flavour related bare $q_i\bar q_j$ states are degenerate, and
which couple to each others through flavour symmetric couplings.
These mesons can then be described  in the ideally mixed reference frame,
i.e. the flavourless states are simply unmixed (like $\ s\bar s,\ c\bar c$). 
Consider now meson loops as shown in  Fig.2 and assume that disconnected
quark-line loop diagrams which violate the OZI rule can be neglected.
To be exact this requires that we have at least 
two degenerate nonets with opposite
charge conjugation, and equal F-like and D-like flavour couplings as described
in more detail in Ref~\cite{NAT4}.
Otherwise singlet states are shifted differently from 
nonsinglet states and  quark-line
disconnected loops (like $\phi \to \omega$ through strange intermediate states) 
do not vanish. 
Here this  $C$-degeneracy is assumed for simplicity, 
but it is relaxed in Ref.~\cite{NAT3} for a model with three flavours,
 where the OZI rule is  broken,  but the isospin subgroup remains unbroken. 

Taking into account these loops, or the 
 vacuum polarization diagrams, the meson spectrum should still be 
consistent with being degenerate, if one disregards possibility of instability. 
This is satisfied, since also the thresholds $m_{ik}+m_{kj}$ 
are degenerate in flavour, and when summed over $k$, each meson 
$ij$ (using the shorthand $q_i\bar q_j\equiv ij$) 
gets an equal contribution from the vacuum polarization 
diagrams. Of course to have this result, not only  the couplings but also 
any  cutoff, $\Lambda$, or subtraction constant involved 
must be independent of flavour. 
Then the  expression for the self energy, which include 
quantum loops can be the same for all $ij$, 
and also the renormalized masses (which now include
"unitarity effects") can remain degenerate. Consequently there is  one very
symmetric  situation, where one knows the solution, 
and our equations must allow for this trivial self-consistent solution. 
{\it But is this solution stable?}
I shall  show that for sufficiently large coupling it is not!

{\it 3. A new mechanism for spontaneous symmetry breaking.}
The simplest,  model is obtained for the case of
only two flavours and scalar mesons coupling to two-body thresholds of  scalar
mesons. Denote the two flavours by 1 and 2. 
Then there are  four mesons, $ij=11 , 12 , 21$
and  $22$, and for each meson there are  meson-meson         
thresholds with different quark content, $m_{ik}+m_{kj},\ k=1,2$.
The inverse propagators, $P^{-1}$, 
get contributions from the meson loops due to 
these thresholds. One finds   a sum of two contributions (cf. Fig. 2)
\begin{equation}
P^{-1}_{ij}(s)= m_0^2- s +
\frac {g^2}{4\pi} \sum_{k=1}^{N_f=2}{\rm F}(s,m_{ik}^2,m_{kj}^2,\Lambda )\ ,
   \label{prop}
\end{equation}
\noindent where $g$ is  the  coupling for each threshold, and $m_0$ is 
the common bare
mass. The function F must have the  unitarity cut, which  is
proportional to two-body phase space.  A simple model for F,
which I  use for the demonstration in this paper, 
has an  imaginary part  given by 
$s\times k/\sqrt s N(s)$, where $k/\sqrt s$ is the usual two-body
phase space factor, and $N(s)$ (in the N/D formalism) is for simplicity,
$N(s)=\theta{\Lambda -k}$ as in an effevtive theory, 
and a real part given by a dispersion relation.
  Denoting as usual the K\"all\'en function 
$(x-y-z)^2-4yz $  by $\lambda(x,y,z)$, 
the imaginary part is thus:
${\rm Im}[{\rm F}] =    
\lambda^{\frac 1 2} (s,m_1^2,m^2_2) \theta (\lambda )\theta
(\Lambda^2-\lambda/(4s) )$.
It vanishes below threshold (the first $\theta$ function) and for
momenta above the cutoff (the second $\theta$ function).
The real part is determined by the dispersion relation:
\begin{equation}
{\rm Re} [{\rm F}(s,m_1^2,m_2^2,\Lambda) ]=   
\frac {1} {\pi} {\cal P}\!\! \int_{0<\lambda<4s'\Lambda^2}\!\!\!\!
 \frac {\lambda^{\frac 1 2}(s',m_1^2,m_2^2)ds' }{s'-s} . \label{regap}
\end{equation}
The  function F can be evaluated analytically and is simply related (up to
subtractions and factors of $s$ in the imaginary part) to the
Chew-Mandelstam, and $H(s)$ functions, which also appear in the literature.   
With this F the coupling constant g is dimensionless. 

The physical meson masses (poles)
are given by the zeroes of the inverse propagators of Eq.(\ref{prop}).
Now eliminate the universal bare mass $m_0$ 
by fixing the mass of one of the mesons, say $m_{11}$.
 One then obtains the following   self-consistency
equations  (cf. Fig.~2):
\begin{equation} 
0 = P_{ij}^{-1}(s=m_{ij}^2) =          \label{twoqeq}
m_{11}^2-s +\frac{g^2}{4\pi} 
\sum_{k=1}^{N_f=2}[{\rm F}( s,m_{ik}^2,m_{kj}^2,\Lambda ) 
-{\rm F}(m_{11}^2,m_{1 k}^2,m_{k1}^2,\Lambda )] \ .\nonumber
\end{equation} 
By construction $m_{11} $ is always the solution of $P_{11}^{-1}(m_{11}^2)$$=0$.
 The two other  masses, $m_{12}$ and $m_{22}$, 
are then determined by the remaining
two equations. From the above discussion
the degenerate solution with equal $m_{ij}=m_{11}$ is one solution.
But this need not be the only solution, and furthermore the symmetric
solution need not be stable. Denoting a small variation from the symmetric
solution in  
the threshold masses  by $\delta_{ij}=m_{ij}^2-m_0^2$, which  
results in a shift $\delta^{\rm out}_{ij}$ in the pole 
positions at the right hand side of Eq. (\ref{twoqeq}), self-consistency
of course requires $\delta^{\rm out}_{ij}=\delta_{ij}$ and stability
$\delta^{\rm out}_{ij}<\delta_{ij}$. I.e., if the latter inequality
is satisfied then starting from some $\delta_{ij}\neq 0$ and
iterating one  converges towards the symmetric solution $\delta^{\rm
out}_{ij}=\delta_{ij}=0$. On the other hand, if 
$\delta^{\rm out}_{ij}>\delta_{ij}$ the symmetric solution is unstable.
Denoting by F$_s=\partial {\rm F}/\partial s|_{s=m_1^2=m_2^2}$
 and by F$_{m^2}=\partial {\rm
F}(s,m_1^2,m^2_2)/\partial m_1^2|_{s=m_1^2=m_2^2}$ 
the self-consistency equations can  be written
\begin{equation}
\delta^{\rm out}_{ij}[N_f \frac{g^2}{4\pi} {\rm F}_s -1]
 +\frac{g^2}{4\pi}{\rm F}_{m^2} \sum_k [\delta_{ik}+\delta_{jk}]=0 \ .
\end{equation}  
This self-consistency requires that small deviations from the symmetric
solution must satisfy the equal spacing rule
\begin{equation}
\delta^{\rm out}_{ij}-\delta^{\rm out}_{ii}=\frac 1 2(\delta^{\rm out}_{jj}-
\delta^{\rm out}_{ii}) \ ,      \label{spacing}
\end{equation}
while the symmetric solution is unstable if
\begin{equation}
r=\frac{\delta^{\rm out}_{ii}} {\delta_{ii}}
= \frac{ N_f {\rm F}_{m^2}} {-N_f {\rm F}_s +4\pi/g^2 }  >  1 
\ . \label{istab2}
\end{equation}

This quantity is plotted in Fig.3 for large values of the coupling constant.
Here $r$ is the slope at origin of the function plotted in Fig. 4. 
Equivalently one can write this condition as a bound on $g^2/(4\pi) $:
\begin{equation}
N_f\frac{g^2}{4\pi} >[{\rm F}_s +{\rm F}_{m^2}]^{-1} \ .
\end{equation}                            
The left hand side is always positive for any reasonable
function F, with correct threshold behaviour. 
Thus  the instability occurs for sufficiently large coupling $g$.
In \cite{NAT4} it was  found that typical coupling constants such as
$g_{\rho\pi\pi}$ and $g_{\sigma \pi\pi}$ very well satisfy this
instability condition, indicating that the instability we discuss actually
occurs in Nature.
 Since this instability depends only on the threshold behaviour of 
relativistic phase space, it 
is a fundamental property related only to the underlying Lorenz invariance
and flavour symmetry of the model.
When $g$ satisfies the condition one must look for another stable solution.
This can only be done numerically since the function F is nonlinear  already
in the simplest possible models. 
To find the stable solution one must solve the nonlinear
equations. This can be done by iterating the Eqs.~(\ref{twoqeq}), starting
with some value off the symmetric solution
(See Fig.~4), in a way analoguous to Fig.~1. 
Using the input masses $m_{ij,in}$  for the threshold masses in the loop one
calculates for which $s=m_{ij,out}^2$ one has zeroes in
 the inverse propagators  Eq.~(\ref{twoqeq}).
The latter are then in the next iteration are used as input masses 
for the thresholds. 
Of course, a  direct way is to solve without
iterations the nonlinear equations on a computer.
These solutions are shown in
Fig.~5a for the functon F defined as above and in the limit when $g$ is  very
large.

{\it 4. The symmetry is broken also in the wave functions.}
How can this unsymmetric solution
arise from a completely flavour symmetric theory? 
 All four mesons are
apart from the "bare $q\bar q$ seed"  composed of clouds of multi-quark
pairs in the form of meson-meson pairs. But these clouds can be 
 different! One can write for the wave functions decompositions:
\begin{equation}
 |ij\!>\propto \sqrt{\frac {g^2}{4\pi}}\bigl [ \sqrt{z^{ij}_1}|i1 ,1j \!>
+\sqrt{z^{ij}_2}|i2 ,2 j \!>\bigr ]+ | ij\! > \ .
\label{wf}
\end{equation}
The coefficients $z^{ij}_k\ $ are generally different and can be computed
from the relative slopes with respect to $s$ 
of the function F evaluated at the stable solution
$z^{ij}_k \propto -\frac{\partial}{\partial s}{\rm F}
(s=m_{ij}^2,m_{ik}^2,m_{kj}^2,\Lambda ) $.
The normalized probabilities $Z^{ij}_1=z^{ij}_1/(z^{ij}_1+z^{ij}_2)$ 
for the stable solution of the model discussed are shown in Fig.~5b.
Only for the unstable, symmetric, solution are the wave functions 
 the same for all $ij$.
The equations (\ref{wf}) are also  self-consistency equations,
i.e., the quantities $ij $ on the right hand side 
stand for a collective multiquark 
state obtained when this quantity is iterated into these
quantities   on the right hand sides. Thus the true 
physical Fock states have components with an arbitrary number of virtual
 quarks or mesons in their wave functions.           
                                     
{\it 5. Concluding remarks.} 
The output stable spectrum is ideally mixed and obeys approximately
the equal spacing rule 
$m_{22}-m_{11}\approx 2(m_{21}-m_{11})$, as one  should expect from Eq. 
(\ref{spacing}).
Thus just as in the real world one can define the constituent
quark mass as approximately 
$M_i\approx m_{ii}/2$, whereby $m_{ij}\approx M_i+M_j$.

A natural very important question which arises in this connection is: 
Where are the Goldstone degrees of freedom and the Goldstone bosons
expected whenever a symmetry is spontaneously broken? In Ref.~\cite{NAT4}
I argue within  a scalar QCD model  that actually the scalar or longitudinal
confined  gluons are the would-be Goldstone bosons, 
not scalar mesons carrying flavour as is usually expected.

In the model presented above there were only S-wave thresholds.  
But the instability increases ($r$ grows) if the threshold involve 
angular momentum factors $k^{2L}$, since this increases the sensitivity
to  the thresholds and in particular $F_{m^2} $  in Eq.~(\ref{istab2}) grows.
Above only scalar mesons were considered. Adding other multiplets of 
different $J^{PC}$ does not change 
the picture  qualitatively, 
since most  multiplets obey  approximately  the  equal spacing rule.
After summing over all thresholds one has 
a similar behaviour in $P^{-1} $.
 Effectively this is approximately equivalent to increasing 
$g$ and $ \Lambda$ in the present model.

The instability occurs for any number of flavours $N_f$.
Of course, how the symmetry group is broken down to a lower symmetry depends
on details of the function F as one 
moves off the symmetry point, and how the mixing between the flavourless
states evolves. In  numerical experiments with $N_f=3$ I find that 
generally an SU2 subgroup remain unbroken (See Ref.~\cite{NAT3,NAT4}.

The assumption of $C$-degeneracy
of multiplets which was made above is certainly not exactly true in reality. 
In the real world this is certainly broken, although
on the average there can be an almost equal contribution from 
F- and D-coupled flavour thresholds. 
Therefore the OZI rule is approximately valid 
for most multiplets.  I relax
the $C$-degeneracy in Ref.~\cite{NAT4}  using a model for 
the ground state mesons, including the pion.  Then  the
OZI rule must be violated by loop diagrams, but the isospin subroup remains
unbroken.

Since  the present  mechanism requires smaller quark masses
it may  be important in the resolving of the strong CP problem, 
for which one resolution would require a vanishing quark mass. Also,
there need be no  contradiction with standard
relations between pseudoscalar masses and usual quark masses~\cite{leut}, 
since the usual quark masses are "measured" through these relations. Including
the  loop effects  discussed 
here, the scale of the quark masses decreases 
while $B$ in $m^2(0^{-+})=$$B(m_{q1}+m_{q2})$ increases.
In lattice gauge theory~\cite{lattice} one also calculates 
light quark masses, but there
one either uses a quenched approximation, or 
make crude approximations for quark loops. In the present
work the loops play a crucial role, without which the mechanism of
 spontaneous breaking would not work. 
Therefore there is no contradiction with present lattice calculations. 

Finally it should be pointed out that this work does revive old ideas from the
60'ies and 70'ies, where one hoped to bootstrap the hadron spectrum through
similar self-consistency equations~\cite{Chew,miya}, but
to my knowledge no demonstration like the one I have presented here was
ever presented. 

I thank especially  Geoffrey F. Chew and Claus Montonen for useful discussions.

\eject
\begin{figure}
\psfig{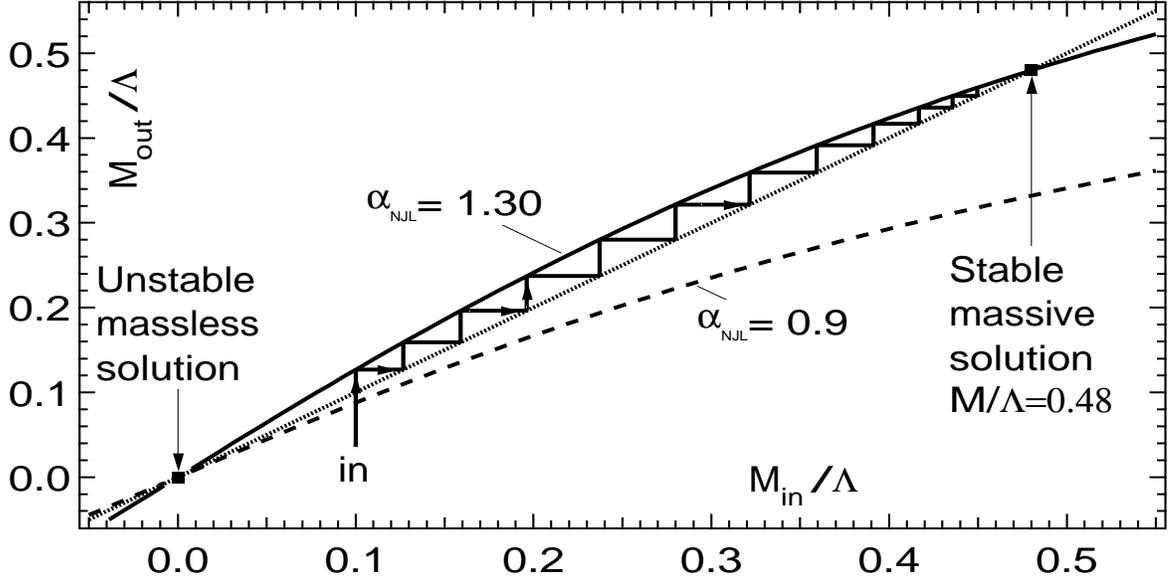}
\caption{Graphical solution of the gap equation. 
For $\alpha_{ NJL} >1$ the slope of the solid curve at the 
origin is $>1$. Then the stable quark mass $M$ is massive, 
otherwise one finds the trivial massless solution (dashed curve).  Here
$\alpha_{ NJL}=1.30$ from which  $M/\Lambda=0.48$, which fits $f_\pi=93$  MeV.
 Then $\Lambda=$0.653 MeV if   $M=M_N/3$.}
\end{figure}                      
\begin{figure}
{\hoffset 0.7cm
\psfig{figure=fig2tornqvist.epsf,width=17cm,height=2.2cm}
}
\vskip 0.3cm
\caption{ The  self-consistency equation diagrammatically. Iterating 
the equation  gives a sum of multiloop diagrams.}
\end{figure}  
\begin{figure}
\label{figgap}
\psfig{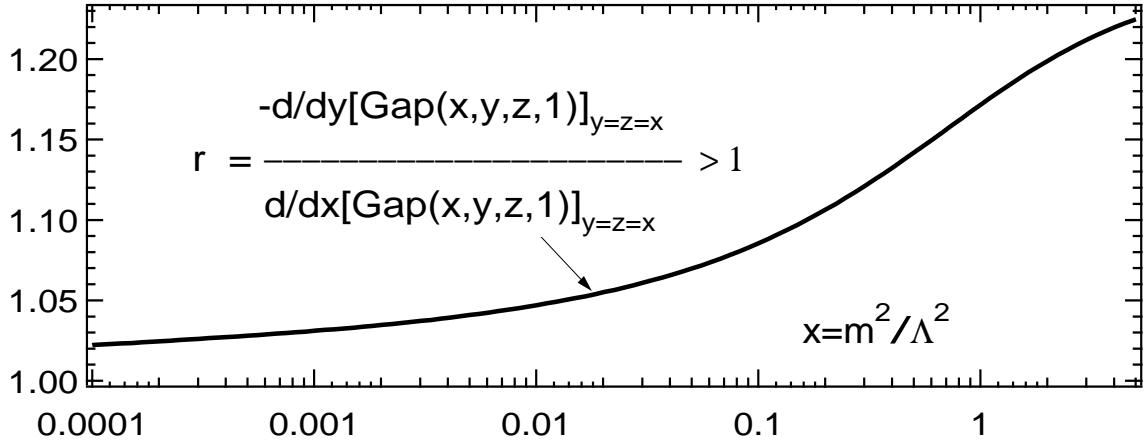}
\caption{ 
The ratio $r$ at the the symmetric point 
$(x=y=z=m_0^2/\Lambda)$ is shown when $g$ is very large.
The fact that $r>1$  impies 
unstability of flavour symmetry for all $\Lambda$ 
and spontaneous flavour symmetry breaking occurs.}
\end{figure}                                      
\begin{figure}
\psfig{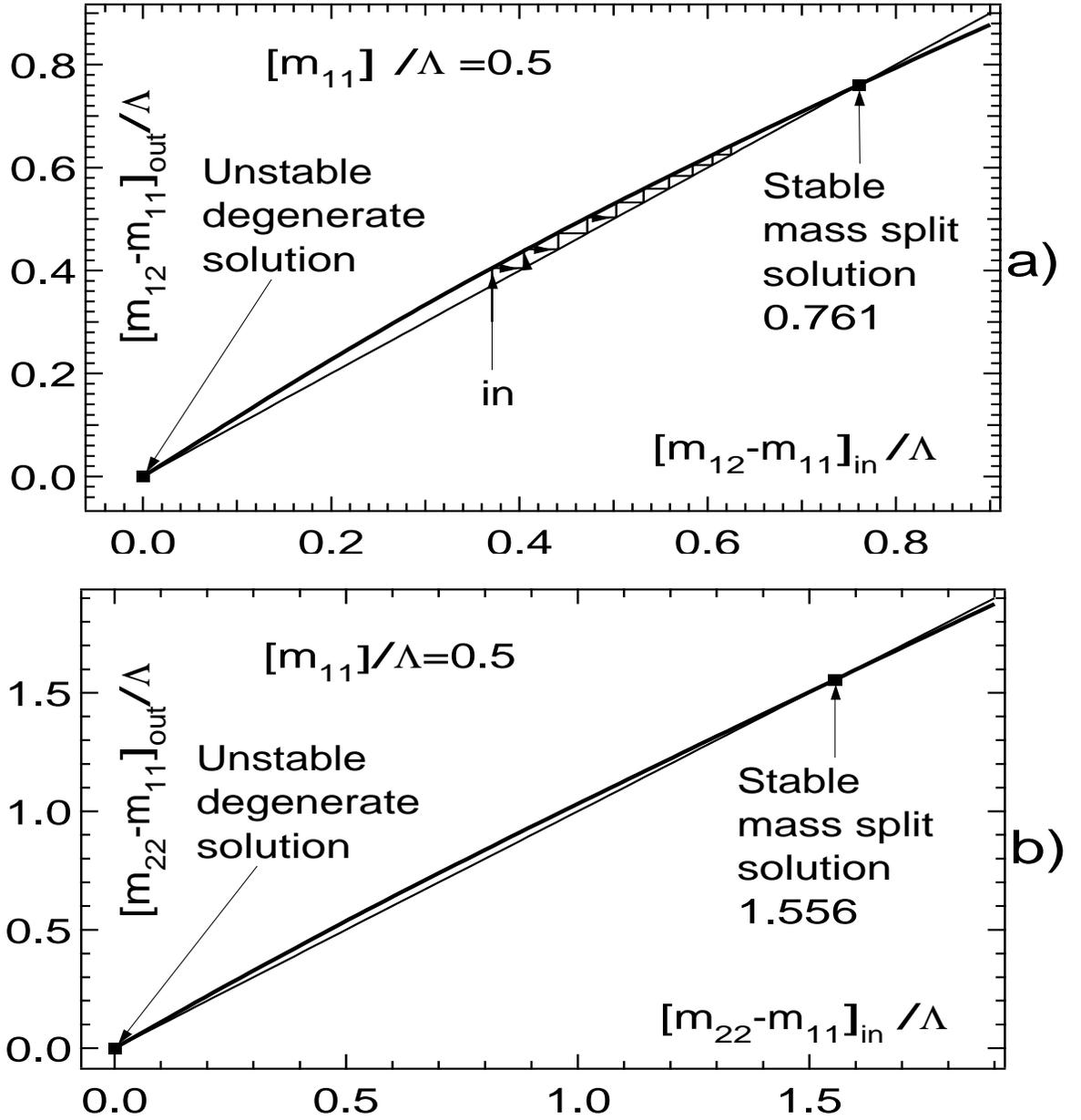}
\label{figinst}
\caption{The spontaneuos breaking of flavour symmetry for two flavours,
when $m_{11}/\Lambda =0.5$ and $g$ very large.
Compare this figure with Fig.~1. The slope
of the curve at the origin is given by  $r$,
which is always $>1$ for all $\Lambda$ (Fig.~3) . 
The symmetric point of equal quark masses is
unstable (like the massless 
point in Fig.~1), while the stable solution has unequal masses.} 
\end{figure}                                  
\begin{figure}
\label{figsol}
\psfig{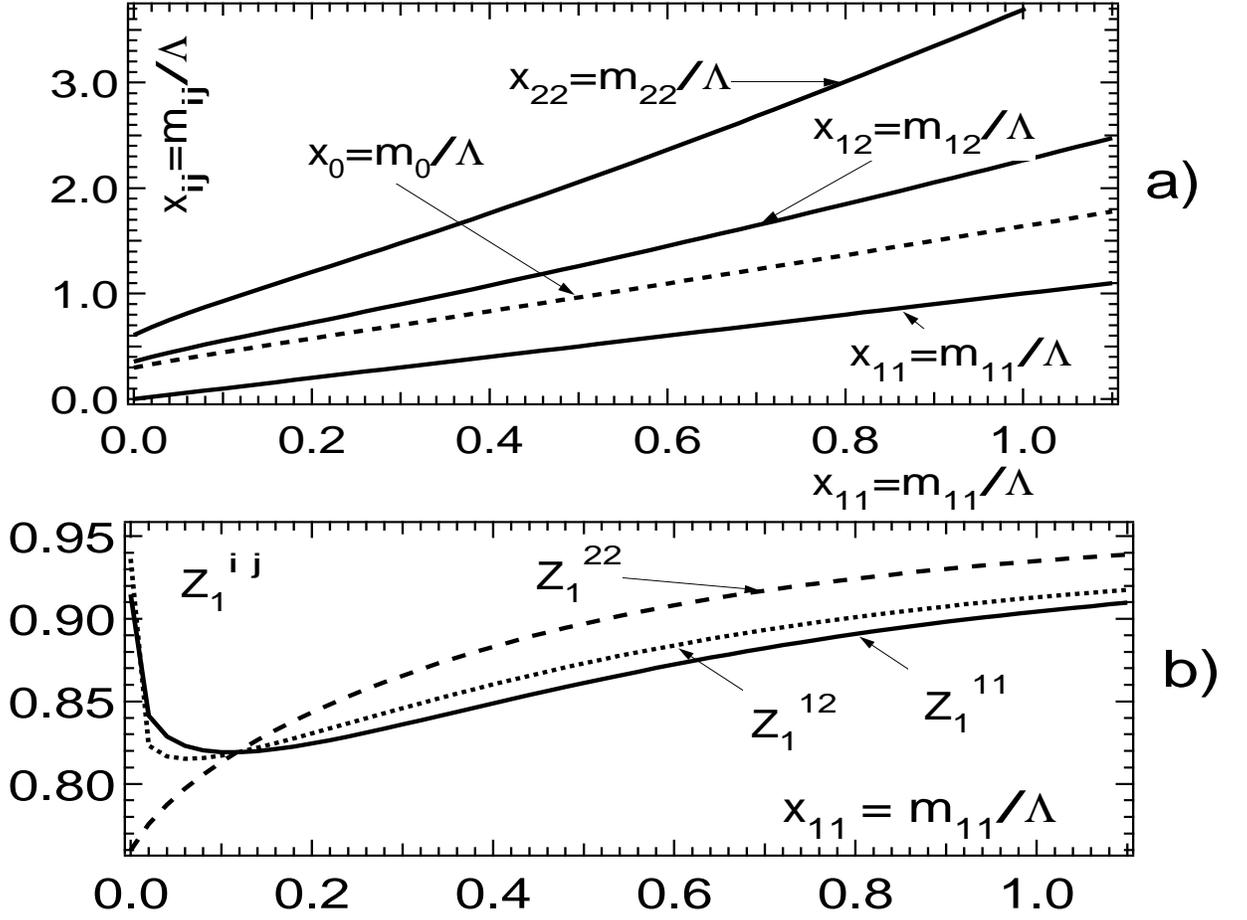}
\caption{a) The spontaneous splitting of meson masses $m_{ij}/\Lambda$ 
 as a function of the lightest meson mass in units of.
Note that the stable solution 
approximately satisfies the equal spacing rule, 
 $m_{22}-m_{12}\approx m_{12}-m_{11}$, in accord with
physical mass splittings. The unstable solution with degenerate masses,
and÷ same subtraction constant. is shown by the dashed curve. 
b) The normalized probabilities $Z^{i,j}_1$  of Eq.~(6). }
\end{figure}
\end{document}